
\input phyzzx  
%
%
\newcount\lemnumber   \lemnumber=0
\newcount\thnumber   \thnumber=0
\newcount\conumber   \conumber=0

\def\myeq{{\rm \chapterlabel.\the\equanumber}}

\def\Lemma{\par\noindent\global\advance\lemnumber by 1
           {\bf Lemma\ (\chapterlabel\the\lemnumber)}}
\def\Corollary{\par\noindent\global\advance\conumber by 1
           {\bf Corollary\ (\chapterlabel\the\conumber)}}
\def\Theorem{\par\noindent\global\advance\thnumber by 1
           {\bf Theorem\ (\chapterlabel\the\thnumber)}}

%
%
\def\e{\adveq\eqno{\rm (\the\equanumber)}}
\def\adveq{\global\advance\equanumber by 1}


%
%
\font\tensl=cmsl10
\font\tenss=cmssq8 scaled\magstep1
\outer\def\quote{
   \begingroup\bigskip\vfill
   \def\endquote{\endgroup\eject}
    \def\par{\ifhmode\/\endgraf\fi}\obeylines
    \tenrm \let\tt=\twelvett
    \baselineskip=10pt \interlinepenalty=1000
    \leftskip=0pt plus 60pc minus \parindent \parfillskip=0pt
     \let\rm=\tenss \let\sl=\tensl \everypar{\sl}}
\def\from#1(#2){\smallskip\noindent\rm--- #1\unskip\enspace(#2)\bigskip}

\def\CALT{\address{Division of Physics, Mathematics
and Astronomy\break
Mail Code 452--48\break
California Institute of Technology\break
Pasadena, CA 91125}}

\def\NIKHEF{\address{NIKHEF-H{~~~~} \break
Kruislaan 409\break
NL--1098 SJ{~~}Amsterdam\break
The Netherlands\break}}

\def\r#1{$\lb \rm#1 \rb$}

%
%
\def\rarrow{\rightarrow}

\def\semidirect{\mathrel{\raise0.04cm\hbox{${\scriptscriptstyle |\!}$
\hskip-0.175cm}\times}}


\def\ref#1{$^{#1}$}

\def\twidle{\tilde}

\def\lb{\lbrack}
\def\rb{\rbrack}

\def\diam{{\hbox{\hskip-0.02in
\raise-0.126in\hbox{$\displaystyle\bigvee$}\hskip-0.241in
\raise0.099in\hbox{ $\displaystyle{\bigwedge}$}}}}
\def\dop{\mathop{{\diam}}\limits}
\def\dw#1#2#3#4{
{\scriptstyle{#4}}\,
{\dop_{#3}^{#1}}
{\scriptstyle{#2}}  }
\def\bw#1#2#3#4#5{{w\left(\matrix{#1&#2\cr#3&#4\cr}\bigg\vert #5\right)}}

\def\sqr#1#2{{\vcenter{\hrule height.#2pt
      \hbox{\vrule width.#2pt height#1pt \kern#1pt
        \vrule width.#2pt}
      \hrule height.#2pt}}}

\def\underwig#1{	
	\setbox0=\hbox{\rm \strut}
	\hbox to 0pt{$#1$\hss} \lower \ht0 \hbox{\rm \char'176}}

\def\bunderwig#1{	
	\setbox0=\hbox{\rm \strut}
	\hbox to 1.5pt{$#1$\hss} \lower 12.8pt
	 \hbox{\seventeenrm \char'176}\hbox to 2pt{\hfil}}
\def\ii{{\rm i}}
\def\ee{{\rm e}}

\def\added{%
\hskip 0pt{\vadjust{\everypar={}
\vtop to 0pt {\hbox{}%
\vskip -13pt\rlap{\hbox to 44pc{\hfil{ \vtop {\hsize=8pc\tolerance=6000%
\hfuzz=.5pc\rightskip=0pt plus 3em\noindent {\tt ADDED}}}}}\vss}}}}%
\def\added{}
\def\changed{%
\hskip 0pt{\vadjust{\everypar={}
\vtop to 0pt {\hbox{}
\vskip -13pt\rlap{\hbox to 44pc{\hfil{ \vtop {\hsize=8pc\tolerance=6000%
\hfuzz=.5pc\rightskip=0pt plus 3em\noindent {\tt CHANGED}}}}}\vss}}}}%
\def\changed{}
%

\frontpageskip=6pt plus 0.5fil minus 2pt

\overfullrule=0pt
\Pubnum={}
\pubtype{CALT 68-1843}
\date{January, 1993}
\titlepage
\title{Braiding in Conformal Field Theory and Solvable Lattice Models}
\author{J\"urgen Fuchs\foot{Heisenberg Fellow}}
\NIKHEF
\author{Doron Gepner\foot{On leave from: Department
of Nuclear Physics,
Weizmann Institute of Science, Rehovot, Israel.}}
\CALT
\vskip2cm
\abstract
Braiding matrices in rational conformal field theory are considered.
The braiding matrices for any two block four point function are computed, in
general, using the holomorphic properties of the blocks and the holomorphic
properties of rational conformal field theory. The braidings of $SU(N)_k$
with the fundamental are evaluated and are used as examples.
Solvable interaction round the face lattice models are constructed from these
braiding matrices, and their Boltzmann weights are given. This allows,
in particular, for
the derivation of the Boltzmann weights of such solvable height models.
\endpage
In a recent publication
\REF\I{D. Gepner, ``Foundations of rational field theory, I'', CALT--68--1825,
November, 1992}
\r\I\ one of the authors has described a universal connection
between rational conformal field theory and solvable lattice models.
The lattice models are built around any rational conformal field
theory, $\cal O$, along with some primary field $x$. The model so obtained,
denoted by IRF$({\cal O},x)$, is described by the partition function,
$$Z=\sum_{\rm configurations} \prod_{\rm faces} \bw a b c d u,\e$$
where $a$, $b$, $c$ and $d$ are four primary fields residing on the
vertices, the Boltzmann weight $\bw a b c d u$ is
defined for each of the faces, $u$ is a spectral parameter which labels
a family of such models, and the product is over all the faces in the
theory. The allowed configurations are limited by the so-called admissibility
condition, which is
$$N_{ax}^b N_{b x}^c N_{cx}^d N_{dx}^a>0,\e$$
where $N_{ab}^c$ is the fusion coefficient of the rational conformal
field theory $\cal O$ (see ref \r\I\ for more details).
Such lattice models have been termed fusion interaction round the face
models, or, in short, fusion IRF. The fusion IRF models are solvable if,
and only if, the Boltzmann weights obey the star-triangle equation (STE), which
implies that the transfer matrices for different spectral parameters, $u$,
are commuting, enabling the exact diagonalization of the transfer matrices.

The problem of finding a solution for the STE is very complicated and only
a limited number of solutions were known, in general. Further, it is not
guaranteed by any means that a given admissibility condition affords such
a solution, and indeed most do not. In ref. \r\I\ a general solution for the
Boltzmann weights of the fusion IRF models was described, satisfying the
STE relation, for any conformal field theory $\cal O$ and any
primary field $x$. The key to the construction of the Boltzmann
weights is using the braiding matrix of the RCFT, which interpolates
two different ways of writing the four point function. The resulting
models automatically satisfy the required STE relation, along with the
precise admissibility condition.

It is noteworthy that all the known solvable lattice models with a second
order phase transition point can, in fact, be written as fusion IRF models,
and thus the above method recovers all the known solvable lattice models,
along with providing a riches of new models. This also allows for the
direct introduction of RCFT methods in the analysis of lattice
models, which is conceptually new.

It is convenient to define the face transfer matrix $X_i(u)$ for a given
lattice model, by the matrix element of the states on the diagonal,
$$\langle a_1,a_2,\ldots,a_i,\ldots,a_n| X_i(u) | a_1,a_2,\ldots,a_i^\prime,
\ldots,a_n\rangle=\bw {a_{i-1}} {a_i^\prime} {a_i} {a_{i+1}} u,\e$$
where the rest of the matrix elements vanish.
The STE then translates to the fact that the face transfer matrices $X_i(u)$
obey the Young--Baxter equation,
$$\eqalign{ X_i(u) X_{i+1}(u+v) X_i(v)&=X_{i+1}(v) X_i(u+v) X_{i+1}(u),\cr
                X_i(u) X_j(v)          &=X_j(v) X_i(u) \qquad{\rm for\ }
|i-j|\geq2.\cr}\e$$
A special case of the Yang--Baxter equation (4) is obtained when both
spectral parameters $u,v$ tend to 
$\ii\infty$. In this case, eqs.\ (4) become
the well known relations of the
braid group. We conclude that $X_i=\lim_{u\rarrow\ii\infty}
X_i(u)$ is a generator of the braid group, i.e., it implements the braiding
of the $i$ and $i+1$ strands in a braid.

Such a braid group representation arises naturally in rational conformal
field theory (for reviews, see for example,
\REF\Frohlich{J. Fr\"ohlich, in: Como 1987 proceedings, Differential Geometric
Methods in Theoretical Physics, p.\ 219
; B. Schroer, same volume, p.\ 289;
J. Fr\"ohlich, in: Cargese Summer Institute 1987, p.\ 71
; L. Alvarez-Gaum\'e, same volume, p.\ 1
}
\REF\Gaume{L. Alvarez-Gaum\'e, C.\ Gomez, and G.\ Sierra, in: The Physics and
  Mathematics of Strings, Knizhnik Memorial Volume, p.\ 16
}
\REF\MSrev{G. Moore and N. Seiberg, Trieste Spring School 1989, p.1}
\r{\Frohlich,\Gaume,\MSrev}).
More precisely, consider the chiral blocks of the
four point function in the $s$ channel,
$${\cal F}_p (z_1,z_2,z_3,z_4)=\langle \psi_i(z_1)\psi_j(z_2)\psi_k(z_3)
\psi_l(z_4) \rangle_p^{},\e$$
where the $\psi_i(z)$ are primary fields in the theory. Here $p$
stands for the primary field exchanged in the $s$ channel. There is a
natural action of the braid group on the conformal blocks by braiding
the points $z_2$ and $z_3$ (or in general any pair of points). This
action is implemented by a finite-dimensional matrix $C$ called the
braiding matrix,
$${\cal F}_p(z_1,z_2,z_3,z_4)=\sum_{p^\prime} C_{p,p^\prime}\left[
\matrix{j&k \cr i & l\cr} \right] {\cal F}_{p^\prime} (z_1,z_3,z_2,z_4),\e$$
where $i$, $j$, $k$ and $l$ label the four primary fields in the correlation
function. Define now, for a fixed primary field $x$, the operator,
  $$\langle a,b,c | X_i |a, d, c\rangle=
C_{b,d} \left[\matrix{ x&x\cr a&c\cr}\right].\e$$
Owing to the fact that $C$ implements the braiding 
of the conformal blocks, it follows that the $X_i$ obey the braid group
relations. Further, the non-vanishing
of the conformal block, eq.\ (5), is precisely equivalent to the admissibility
condition eq.\ (2). We conclude that $X_i$ is a valid fusion IRF face transfer
matrix. Albeit, we do not have any spectral parameter in it. Thus we need
to construct a solution $X_i(u)$ such that $X_i=\lim_{u\rarrow \ii\infty}
X_i(u)$, and $X_i(u)$ obey the 
star-triangle equation.
This can be done in a universal,
and essentially unique, way, as described in ref.\ \r\I. Here we shall
make the simplifying assumption that the field $x$ is a `fundamental' field,
i.e., we shall assume that the operator products $x\cdot x$ and $x\cdot
\bar x$ have precisely two primary fields in each. In this case the
face transfer matrices assume a particularly simple form,
$$X_i(u)=\sin(\lambda-u) + \sin u H_i,\e$$
where $\lambda=\pi(\Delta_1-\Delta_2)/2$ is the crossing parameter,
and $\Delta_i$ are the conformal
dimensions of the fields in the product $x\cdot x$. The $H_i$
are defined as
$$H_i=\ee^{-\ii\lambda}-X_i,\e$$
and obey the $A$-type Hecke algebra relations,
$$\eqalign{
H_i H_{i+1} H_i-H_i&=H_{i+1} H_i H_{i+1}-H_{i+1},\cr
H_i H_j&=H_jH_i \qquad{\rm for \ }|i-j|\geq2,\cr
H_i^2&=(2\cos\lambda) H_i,\cr}\e$$
{}from which the STE can be readily verified.

Our purpose in this note is to explicitly compute the braiding matrices
appearing in rational conformal field theory. This will then provide
concrete expressions for the Boltzmann weights of the fusion IRF models.
Using the holomorphic properties of the conformal blocks, we will derive a
general result for the braiding of any fundamental field, in any rational
conformal field theory, expressing the result in terms of conformal
dimensions only. The $SU(N)_k$ height models will be constructed for the
purpose of illustration.

Let us, at first, demonstrate the technique by specializing to the case
of IRF(SU$(N)_k,N,N)\equiv\,$IRF$(SU(N)_k,[N])$.
Here $SU(N)_k$ denotes the corresponding
current algebra, or equivalently the corresponding WZNW model. Also,
$[N]$ stands for the fundamental representation of $SU(N)$, of
highest weight $\Lambda_{[N]}=\Lambda_{(1)}$
($\Lambda_{(i)}, \,i = 1, ... , N - 1$, denote the
fundamental weights of $SU(N)$); this field has indeed
only two primaries in its operator product, $[N]\cdot[N]=\psi_1+\psi_2$,
where $\psi_i$ refer to the representations with the highest weight
$2\Lambda_{(1)}$ and $\Lambda_{(2)}$, respectively.
We want to derive the braiding matrix $B$ by making use of the fact that
the analytic properties of the correlation functions of conformal field
theory are to a large extent determined by their singularities.  Our
main task is to analyze the transformation property of the four point
function
$${\cal G}(z)=\langle\Lambda_a (z) \Lambda_{(1)} (0) \Lambda_{(1)}
(1) \Lambda_b (\infty)\rangle \e$$
under the substitution $z \mapsto 1 - z$.  In eq.\ (11), $\Lambda$ is used as
a short-hand notation for the WZNW primary field $\phi^\Lambda$, which
carries the irreducible highest weight module with highest weight
$\Lambda$.

We are interested in the case where precisely two chiral blocks
contribute to ${\cal G}$. This means that the $SU(N)$ tensor product
$$\Lambda_a \times \Lambda_{(1)} \times \Lambda_{(1)} \times
\Lambda_b \e$$
must contain precisely two singlets.  To be able to describe
conveniently a necessary and sufficient condition for this situation, we
introduce some notation.  First, we write the weight $\Lambda_a$ in the
form
$$\Lambda_a = \sum_{j = 1}^{n_{a}} \Lambda_{(a_{j})} \e$$
where the labels $a_j$ are ordered such that $a_j \leq a_{j + 1}$ for
all $j = 1, ... , n_a - 1$, and analogously for $\Lambda_b$.  Next, we
define $\Lambda_{(N)}: = \Lambda_{(0)} \equiv 0$, and allow for $1 \leq
a_j \leq N$ so that without any loss of generality, it can be assumed that
the number of terms appearing in eq.\ (13), and in the corresponding
decomposition of $\Lambda_b$ are identical, $n_a = n_b$.

The condition can now be stated as follows: the tensor product
(12) contains exactly two singlets iff the weights
$\Lambda_a$ and $\Lambda_b$ are related by
$$(\Lambda_b)^* = \Lambda_{a;l,m} : = \Lambda_a - \Lambda_{(a_{l})} +
\Lambda_{(a_{l} + 1)} - \Lambda_{(a_{m})} + \Lambda_{(a_{m} + 1)} \e
$$
for some integers $l$ and $m$, satisfying $1 \leq l, m \leq n_a$ and $l
\not= m$ (in particular, the weight defined this way must be dominant,
which implies $a_l < a_{l + 1}$ and $a_m < a_{m + 1}$).
If this condition is fulfilled, the families exchanged in the $s$ and
$t$ channels (i.e., those which give rise to the chiral blocks for $z
\simeq 0$ and $z \simeq 1$, respectively) correspond to the weights
$$\Lambda_1^{(s)}=\Lambda_1^{(t)}=\Lambda_{a;l}
\quad {\rm and} \quad \Lambda_2^{(s)}=\Lambda_2^{(t)}=\Lambda_{a;m}
\,\, , \e$$
respectively, while in the $u$ channel $(z \simeq \infty)$ the weights
are, of course, $\Lambda_{(2)}$ and $2\Lambda_{(1)}$.
In eq.\ (15), we introduced the notation
$$\Lambda_{a;l}:=\Lambda_a - \Lambda_{(a_{l})}+
\Lambda_{(a_{l} + 1)} \e$$
for $0\leq l<N$.

The exponents, i.e., the leading singularities of the chiral blocks of
${\cal G}$, can  be expressed in terms of the conformal dimensions
$\Delta (\Lambda)$ of the various primary fields encountered above; they
read
$$\eqalign{
{\rm at}~ z = 0 :\qquad\qquad \alpha_1^{(0)} &= - \Delta
(\Lambda_a)-\Delta (\Lambda_{(1)})+\Delta(\Lambda_{a;l}),\cr
\alpha_2^{(0)}&=-\Delta(\Lambda_a)-\Delta (\Lambda_{(1)})+
\Delta(\Lambda_{a;m}) + 1;\cr
&\cr
{\rm at}~ z = 1 :\qquad\qquad\alpha_1^{(1)}&=- \Delta (\Lambda_a)-\Delta
(\Lambda_{(1)})+\Delta (\Lambda_{a;l}),\cr
\alpha_2^{(1)}&=-\Delta(\Lambda_a)-\Delta(\Lambda_{(1)})+
\Delta(\Lambda_{a;m});\cr
&\cr
{\rm at}~ z=\infty :\quad\qquad\alpha_1^{(\infty)}&=\Delta (\Lambda_a)-
\Delta
(\Lambda_{a;l,m})+\Delta (\Lambda_{(2)}),\cr
\alpha_2^{(\infty)}&=\Delta (\Lambda_a)-\Delta (\Lambda_{a;l,m})+
\Delta(2\Lambda_{(1)}).\cr}\e$$
The addition of unity to $\alpha_2^{(0)}$ as compared with
$\alpha_2^{(1)}$ corresponds
\REF\Fuchs{J. Fuchs, Nucl. Phys. B 328 (1989) 585}\r\Fuchs\
to the fact that, with the natural
definition of chiral blocks appropriate for $z$ close to a fixed
singular point $z_0$, only one of the leading singularities of a WZNW
correlation function at $z_0$ can come from the relevant primary field,
whereas any other leading singularity must come from a descendant
field.  In eq.\ (17) we have chosen $z_0 = 0$ as the distinguished singular
point, corresponding to the blocks at $z \simeq 0$; below we will also
have to consider $z_0 = 1$ which is tantamount to replacing
$\alpha_2^{(0)}$ and $\alpha_2^{(1)}$ by $\alpha_2^{(0)} - 1$ and
$\alpha_2^{(1)} + 1$, respectively.

The conformal dimensions of WZNW primary fields are given by $\Delta
(\Lambda) = \kappa {\cal C}(\Lambda)$ with $\kappa = {1\over k+N}$,
where ${\cal C}$ is the eigenvalue of the quadratic Casimir operator.  In the
notation used in eq.\ (13), the Casimir eigenvalue is
$$	{\cal C} (\Lambda_a) = {1\over 2} \left[ \sum_{j = 1}^{n_{a}} a_j
(N - a_j + 2n_a - 2j + 1) - {1\over N} (\sum_{j = 1}^{n_{a}} a_j)^2\right]
\,\, . \e$$
{}From this formula we deduce, in particular,
$$	{\cal C} (\Lambda_{a;l}) = {\cal C} (\Lambda_a) - a_l - l + n_a
+ {1\over 2} (N - {1\over N}) - {1\over N} \sum_{j = 1}^{n_{a}} a_j
\e$$
and,
$$	{\cal C} (\Lambda_b) \equiv {\cal C} (\Lambda_{a;l,m}) = {\cal
C}(\Lambda_a) - a_l - a_m - l - m + 2n_a + N - {2\over N} (1 + \sum_{j =
1}^{n_{a}} a_j) \e$$
if $\Lambda_{a;l}$ and $\Lambda_b$ are of the form eq.\ (16), respectively
eq.\ (14). 
Inserting this result into eq.\ (17), one  verifies that
$$	\mathrel{\mathop{\sum_{\scriptstyle {i = 1,2}
\atop\scriptstyle{j = 0,1,\infty}}}} \alpha_i^{(j)} = 1 \,\, , \e
$$
a relation that indeed must be satisfied
\REF\Blok{B. Blok and S. Yankielowicz, Nucl. Phys. B 321 (1989) 717}
\r{\Blok,\Fuchs}\
by the exponents.

Any conformal field theory correlation function with two chiral blocks
satisfies a second order linear differential equation.  It can be shown
\REF\Fuchstwo{J. Fuchs, Nucl. Phys. B 386 (1992) 343}\r\Fuchstwo\
that for WZNW correlators, this differential equation does not possess
the so-called apparent singularities; as a consequence
\r{\Blok,\Fuchs}, the chiral
blocks are simple combinations of powers and hypergeometric functions,
with the parameters determined uniquely by the exponents eq.\ (17).  In the
present situation, we find that (up to an overall factor $[z (1 -
z)]^{\alpha_{1}^{(0)}}$ which is irrelevant for the considerations
below, and which we suppress in the sequel)
$$	{\cal G}_1 (z) \propto~_2\!F_1 (\gamma, \delta; 1 - \alpha; z)
 \,\, ,\qquad~~ \qquad$$
$${\cal G}_2 (z) \propto z^\alpha~_2\!F_1 (\alpha + \gamma, \alpha +
\delta; 1 + \alpha; z) \,\, . \e$$
Here $_2\!F_1 (a,b; c;z) \equiv \sum_{j = 0}^{\infty} [\Gamma (a + j)
\Gamma(b + j) \Gamma(c)/j!\Gamma (a) \Gamma (b) \Gamma (c + j)]z^j$
is the hypergeometric series.  The parameters $\alpha, \gamma$ and
$\delta$ are given by
$$	\alpha = \alpha_2^{(0)} - \alpha_1^{(0)} = 1 + {1\over k + N}
\,a_{lm}, \e$$  
with
$$	a_{lm} : = a_l - a_m + l - m \,\, , \e$$
and by
$$	\matrix{\gamma &= \alpha_1^{(0)} + \alpha_1^{(1)} + \alpha_1^{(\infty)} =
1 - \alpha - {1\over k + N} \,\, ,\cr
\delta &= \alpha_1^{(0)} + \alpha_1^{(1)} + \alpha_2^{(\infty)} = 1 -
\alpha + {1\over k + N} \,\, .\cr}\e$$
In particular, the blocks can be expressed in terms of $\alpha$ (and
hence $a_{lm}$) alone,
$$	\matrix{{\cal G}_1 (z) &= {}_2\!F_1 (1 - \alpha - {1\over k + N}, 1
- \alpha + {1\over k + N} ; 1 - \alpha; z) \, ,~~~ \cr
{\cal G}_2 (z) &= {1\over k + N} {\cal N}_0 z^\alpha\cdot {}_2\!F_1
(1 - {1\over k + N}, 1 + {1\over k +N}; 1 + \alpha; z) \,\, . \cr} \e$$
Here we defined
$$	{\cal N} _0 = - \ee^{\pi \ii \alpha} {\Gamma (1 - \alpha) \Gamma
(\alpha + {1\over k + N})\over \Gamma (\alpha + 1) \Gamma (1 - \alpha
+ {1\over k + N})} \sqrt{{s (\alpha + {1\over k + N})\over s (\alpha -
{1\over k + N})}} \e$$
and
$$	s (x) := \sin (\pi x) \,\, , \e$$
whereby we fixed both the overall normalization, as well as the relative
normalization of the two blocks (which
affects the normalization of operator product coefficients) in a manner
such as to simplify some of the formulae below.

As noted after eq.\ (17), the functions eq.\ (26) are the blocks
appropriate for $z \simeq 0$; for $z \simeq 1$ we have to replace
$\alpha$ by $\alpha - 1$, leading to the blocks
$$\eqalign{\tilde{\cal G}_1 (z)&= {1\over (k + N) (\alpha - 1)}
{\cal N}_1 \cdot{}_2\!F_1 (1 - \alpha - {1\over k + N}, 1 - \alpha +
{1\over k + N}; 2 - \alpha; z) \, , \cr
\tilde{\cal G}_2 (z)&= \alpha {\cal N}_0 {\cal N}_2 z^{\alpha - 1}
\cdot {}_2\!F_1 (- {1\over k + N}, {1\over k +N}; \alpha; z).
\cr} \e$$
Here we again introduced some normalization constants, ${\cal N}_1$
and ${\cal N}_2$,
but this time we
are not free to fix them at will.  Rather, the normalizations are determined
uniquely by the condition
(33) below.  At the present stage, however, ${\cal N}_1$ and ${\cal
N}_2$ cannot yet be determined, inasmuch as we have only used properties of
{\it linear} differential equations.

We have now gathered all the information needed for the construction of
the braiding matrix.  This is the matrix $B$ which relates the blocks
${\cal G}_i (1 - z)$ to the blocks $\tilde{\cal G}_i (z)$.  To obtain
$B$, we only need to apply the standard formula expressing the
hypergeometric function at $1 - z$ in terms of hypergeometric functions
at $z$ (see, e.g.,
\REF\Erd{A. Erd\'elyi, ed., {\it Higher Transcendental Functions}
vol. I (McGraw-Hill, New York 1953)}\r{\Erd, p.108}).
 Using also functional identities of
the Gamma function, in particular $\Gamma (x) \Gamma (1 - x) = \pi/s
(x)$, we can write the result in the form
$$	{\cal G}_i (1 - z) = \sum_{j = 1}^2 B_{ij} \tilde{\cal G}_j (z)
\e$$
for $i = 1,2,$ with the two by two matrix
$$	B = {s({1\over k+N})\over s(\alpha)} \pmatrix{- {\cal N}_1^{-1}
& - {\cal N}_2^{-1} \ee^{-\pi \ii\alpha} \rho\cr
- {\cal N}_1^{-1} \ee^{\pi \ii\alpha} \rho & {\cal N}_2^{-1}\cr} \,\, ,
\e$$
where
$$	\rho = {1\over s({1\over k + N})} \sqrt{s (\alpha + {1\over k +
N})\, s (\alpha - {1\over k + N})} \,\, . \e$$

It remains to fix the normalization constants ${\cal
N}_1$ and ${\cal N}_2$.  This is done using the close relationship
between braiding and monodromy.  Namely, $B$ must satisfy
\REF\Reh{K.-H. Rehren and B. Schroer, Nucl. Phys. B 312 (1989) 715}
\r\Reh
$$(DB)^2 = {\bf 1} \,\, , \e$$
where $D$ is the diagonal matrix
$$D = \exp [ - \pi \ii (\Delta (\Lambda_a) + \Delta (\Lambda_b))]
\pmatrix{\exp [2 \pi \ii \Delta (\Lambda_{a;l})] & 0 \cr
0 & \exp[2 \pi \ii \Delta (\Lambda_{a;m})] \cr}\,\, , \e$$
i.e.,
$$D=- \ee^{\pi \ii/N (k + N)} \pmatrix{\ee^{- \pi \ii\alpha} & 0\cr
0 & \ee^{ \pi \ii \alpha}\cr} \,\, . \e$$

Inserting eq.\ (34), along with eq.\ (31), into the relation eq.\ (33),
we conclude that ${\cal N}_1=\exp \left[{\pi \ii\over N(k+N)}
-\pi\ii\alpha\right]$ and ${\cal N}_2
=\exp \left[{\pi \ii\over N(k+N)}+\pi\ii\alpha\right]$.  Hence
$$B=-s ({1\over k + N}) s^{-1} (\alpha) \,\ee^{-\pi \ii/N (k + N)}
\pmatrix{ \ee^{ \pi \ii\alpha} &   \rho\cr
 \rho & - \ee^{- \pi \ii\alpha}\cr}\,\, . \e$$
This result can be expressed in terms of the quantity $q:=
\ee^{2\pi \ii/(k + N)}$ and the function
$$[x] : = {q^{x/2} - q^{-x/2}\over q^{1/2} - q^{-1/2}} = {s
({x\over k + N})\over s ({1\over k + N})} \,\, . \e$$
Eq.\ (36) then assumes the form
$$B = {q^{-1/2N}\over [a_{lm}]} \pmatrix{q^{ a_{lm}/2} &
\sqrt{[a_{lm} + 1] [a_{lm} - 1]}\cr
\sqrt{[a_{lm} + 1] [a_{lm} - 1]} & -q^{- a_{lm}/2}\cr}\,\, . \e$$

Note that for the special case of $SU(2)$, $N=2$, the braiding matrix
agrees precisely with the one computed earlier
\REF\TSUK{A. Tsuchiya and Y.\ Kanie, Adv. Studies in Pure Math.
16 (1988) 297}
\r\TSUK.
As in the
case of $SU(2)$, we can now proceed to construct the lattice model
IRF(SU$(N)_k,N,N)$ using the formalism of section 7 of \r\I.  The crossing
parameter is given by (compare eq.\ (7.27) of \r\I)
$\lambda=\pi(\Delta_1-\Delta_2)/2={\pi\over k+N}$, where $\Delta_1$ and
$\Delta_2$ are the dimensions of the two fields exchanged in this channel.
The generator of the Hecke algebra is, as usual,
$$H=q^{-1/2}-\ee^{-\pi\ii(\Delta_1+\Delta_2-2\Delta(\Lambda_{(1)}))}B,\e$$
and it obeys the usual Hecke relation,
$$H^2=\beta H,\qquad {\rm where\ } \beta=2\cos\lambda.\e$$
The Hecke algebra elements are
$$H_i=\dw \lambda {\lambda+e_k} {\lambda+e_j+e_l} {\lambda+e_j} \,=
(1-\delta_{jl}) {\left[s_{jl}(\lambda+e_j)\, s_{jl}(\lambda+
e_k) \right]^{1\over2}\over s_{jl}(\lambda)},\e$$
where $s_{jl}(\lambda)=\sin[({\pi\over N})\,
(e_j-e_l)\cdot \lambda]$, and where $e_j=\Lambda_{(a_j)}-\Lambda_{(a_{j+1})}$.
This expression corresponds to a well known representation of the Hecke algebra
\REF\Wenzl{H. Wenzl, Representations of Hecke algebras and subfactors, Univ. of
Pennsylvania Thesis (1985), Invent. Math. 92 (1988) 349}\r\Wenzl.
{}From this representation we find, by substituting it into eq.\ (8),
the Boltzmann weights of the trigonometric lattice IRF model, which are,
$$\eqalign
{\dw a {a+e_l} {a+2e_l} {a+e_l}&=[1-u],\cr
\dw a {a+e_l} {a+e_l+e_m} {a+e_l}&={[a_{lm}+u]\over [a_{lm}]},\cr
\dw a {a+e_m} {a+e_l+e_m} {a+e_l}&=p\,[u] {\sqrt{[a_{lm}+1][a_{lm}-1]}
\over [a_{lm}] },\cr}\e$$
where $p=\pm1$ corresponds to two different solutions (given by $B$ or its
complex conjugate matrix, which is \r\I\ the braiding matrix of $SU(N)_{-1}$).
Surprisingly, these are precisely the Boltzmann weights of the $SU(N)$
models described in ref.
\REF\Jimbo{M. Jimbo, T. Miwa and M. Okado, Lett. Math. Phys. 14 (1987) 123}
\r\Jimbo,
at the critical limit, $p=0$.  This fully illustrates
the connection
between IRF models and RCFT, giving, in this particular
instance, the Boltzmann weights of the solvable lattice models
IRF(SU$(N)_k,N,N)$.

One can easily extend the results described in this note to other
modular invariants of $SU(N)$, and their extended algebras. It is known
that the same blocks appear in all modular invariants, and that the
problem of writing the braiding matrices is a simple sesqui-linear
re-juxtapositioning of the conformal blocks
(see, for example,
\REF\Trivedi{M.R.  Douglas and S.P. Trivedi, Nucl. Phys. B320 (1989) 461}
\REF\Jur{J. Fuchs, Phys. Rev. Lett. 62 (1989) 1705;
J. Fuchs and A. Klemm, Ann. Phys. 194 (1989) 303}
\r{\Trivedi,\Jur}).
Substituting the so obtained braiding matrices into eq.\ (8)
we would find  new solvable IRF models, and
an explicit solution for their Boltzmann weights. In fact, the Boltzmann
weights
of such models are yet to be explored.
Another approach for the construction of lattice models, based on graph theory,
has been advocated by Pasquier (the D-E cases)
\REF\Pas{V. Pasquier, J. Phys. A20 (1987) L217}
\r\Pas,
and generalized to all $SU(N)$ by di Francesco and Zuber
\REF\Zuber{P. di Francesco and J.B. Zuber, Nucl. Phys. B338 (1990) 602}
\r{\Zuber}. Though the two approaches are bound to be related, the
precise connection requires more study. Many of the other modular invariants
can be related to quotients (`orbifolds') of the corresponding conformal field
theories \REF\Gepwit{D. Gepner and E. Witten, Nucl. Phys. B278 (1986) 493}
\r\Gepwit. The IRF models built on these should
correspond to the quotient procedure of the IRF models
\REF\Kostov{I. Kostov, Nucl. Phys. B300 [FS22] (1988) 559}
\REF\Fen{P. Fendley and P. Ginsparg, Nucl. Phys. B324 (1989) 549}
\REF\Fentwo{P. Fendley, J. Phys. A22 (1989) 4633}
\r{\Kostov,\Fen,\Fentwo}. More generally, taking first a quotient of
an arbitrary rational conformal field theory and then using equation (8) is
equivalent to the IRF quotient procedure as applied to the original
theory. Note, however,
that taking $x$ to be a twisted field give rise to an entirely new IRF model,
which is not a quotient of the original IRF.

\def\s{\mathop{\rm s}\nolimits}
\def\cG{{\cal G}}
\def\cN{{\cal N}}
Let us turn now to the generalization of the above calculation of the braiding
matrix for any two block correlation function. The calculation is a
straightforward extension of the one above, where we assume that there are
no apparent singularities.

Consider the correlation function
$$\cG(z)=\langle \phi(z)\phi_s(0) \phi_t(1) \phi_u(\infty)\rangle,\e$$
where $\phi$, $\phi_s$, $\phi_t$ and $\phi_u$ are four arbitrary primary
fields.
Further, assume that this correlation function corresponds to two blocks, i.e.,
it receives contributions from exactly two fields in each channel,
$$\phi\cdot \phi_a=\phi_1^{(a)}+\phi_2^{(a)},\e$$
where $a=s$, $t$, or $u$ label the channel.
The exponents of the correlator (43) 
are expressed in terms of the conformal dimensions,
$$\eqalign{\alpha_i^{(0)}&=-\Delta-\Delta_s+\Delta_i^{(s)}+i-1,\cr
           \alpha_i^{(1)}&=-\Delta-\Delta_t+\Delta_i^{(t)},\cr
           \alpha_i^{(\infty)}&=\Delta-\Delta_u+\Delta_i^{(u)},\cr}\e$$
for $i=1,2$, 
where $\Delta=\Delta(\phi)$,
$\Delta_a=\Delta (\phi_a)$, and $\Delta_i^{(a)}=\Delta (\phi_i^{(a)})$.
The conformal dimensions appearing here are restricted by the exponent sum rule
$$\sum_{i=1,2\atop j=0,1,\infty} \alpha_i^{(j)}=1.\e$$

The exponents describe the order of the singularities 
of the correlation function
$\cG(z)$ at the points $z=0$, $z=1$ and $z=\infty$. Along with the
holomorphicity this implies that $\cG$ is given by hypergeometric functions
(suppressing again an overall power factor),
$$\eqalign{
  \cG_1(z)&={}_2\!F_1(\gamma,\delta;1-\alpha;z),\cr \cG_2(z)&=
  \cN_0 z^\alpha \cdot {}_2\!F_1(\alpha+\gamma,\alpha+\delta;1+\alpha;z),
\cr}\e$$
which are the blocks relevant in $z$. The blocks relevant in $1-z$ are
\def\ctG{{\twidle{\cal G}}}
$$\eqalign{
\ctG_1(z)&=\cN_1\cdot {}_2F_1(\gamma,\delta;
1-\beta;z),\cr
\ctG_2(z)&=\cN_0\cN_2 z^\beta\cdot {}_2 F_1(\beta+\gamma,\beta+\delta
;1+\beta;z).\cr}\e$$
Here the parameters $\alpha$, $\beta$, $\gamma$ and $\delta$ are
$$\eqalign{\alpha&=\alpha_2^{(0)}-\alpha_1^{(0)}=\Delta_2^{(s)}
  -\Delta_1^{(s)} +1,\cr
           \beta&=\alpha_2^{(1)}-\alpha_1^{(1)}=\Delta_2^{(t)}-
\Delta_1^{(t)},\cr
           \gamma&=\alpha_1^{(0)}+\alpha_1^{(1)}+\alpha_1^{(\infty)}=
   \Delta_1^{(s)}+\Delta_1^{(t)}+\Delta_1^{(u)}-\Delta-\Delta_s-\Delta_t-
      \Delta_u,\cr
\delta&=\alpha_1^{(0)}+\alpha_1^{(1)}+\alpha_2^{(\infty)}=\Delta_1^{(s)}
+\Delta_1^{(t)}+\Delta_2^{(u)}-\Delta-\Delta_s-\Delta_t-\Delta_u.\cr}\e$$
{}From the sum rule (46)
, the parameters satisfy
$$\alpha+\beta+\gamma+\delta=1.\e$$
It is readily verified that the blocks obey the braiding property,
$$\cG_i(1-z)=\sum_{j=1}^2 B_{ij} \ctG_j(z),\e$$
where the braiding matrix is
$$B=\pmatrix{\cN_1^{-1} {\Gamma(1-\alpha)\Gamma(\beta)\over
     \Gamma(\beta+\gamma)\Gamma(\beta+\delta)} &
 \cN_0^{-1} \cN_2^{-1} {\Gamma(1-\alpha) \Gamma(-\beta)\over
       \Gamma(\gamma)\Gamma(\delta)} \cr
\cN_0 \cN_1^{-1} {\Gamma(1+\alpha) \Gamma(\beta) \over
     \Gamma(1-\gamma) \Gamma(1-\delta)} &
\cN_2^{-1} {\Gamma(1+\alpha) \Gamma(-\beta) \over \Gamma(\alpha+\gamma)
   \Gamma(\alpha+\delta)}\cr}.\e$$

The normalizations $\cN_i$ need to be determined. This is done using the
matrix equation (33), i.e. $(DB)^2={\bf 1},$
where the diagonal matrix $D$ is given by
  \added
$$D=\pmatrix{d_1&0 \cr 0&d_2} = \exp[-\pi\ii (\Delta+ \Delta_u)]
    \pmatrix{ \exp[2\pi\ii\Delta_1^{(s)}] & 0 \cr 0 &
  \exp[2\pi\ii\Delta_2^{(s)}]  } . \e$$
{}From the off diagonal elements of this constraint we find
$$\cN_2={d_2\over d_1} {\Gamma(\alpha)\Gamma(-\beta)\Gamma(\beta+\gamma)
  \Gamma(\beta+\delta) \over
         \Gamma(-\alpha)\Gamma(\beta)\Gamma(\alpha+\gamma)
              \Gamma(\alpha+\delta)}\, \cN_1.\e$$
For the diagonal elements we then obtain 
$$((DB)^2)_{11}^{}=((DB)^2)_{22}^{}=(d_1\cN_1^{-1})^2 \,
{ \Gamma(1-\alpha) \Gamma(\beta)
\Gamma(\alpha+\gamma)\Gamma(\alpha+\delta) \over
    \Gamma(\alpha)\Gamma(1-\beta)\Gamma(\beta+\gamma)\Gamma(\beta+\delta)}.\e$$
Thus,
$$\eqalign{
\cN_1&=d_1\sqrt{\Gamma(1-\alpha)\Gamma(\beta)\Gamma(\alpha+\gamma)
    \Gamma(\alpha+\delta) \over
        \Gamma(\alpha)\Gamma(1-\beta) \Gamma(\beta+\gamma)
                \Gamma(\beta+\delta) },\cr
  \cN_2&=d_2\sqrt{\Gamma(1+\alpha)\Gamma(-\beta)\Gamma(\beta+\gamma)
          \Gamma(\beta+\delta) \over \Gamma(-\alpha) \Gamma(1+\beta)
\Gamma(\alpha+\gamma)\Gamma(\alpha+\delta)}.\cr}\e$$
Inserting the normalizations into $B$ one finds, in particular,
$$d_1 B_{11}=d_2 B_{22}=\sqrt{s(\beta+\gamma) s (\beta+\delta)\over
   s(\alpha) s(\beta)},\e$$
   \added %
i.e.\ the diagonal elements of $B$ can be rewritten purely in terms
of phases and $q$-integers. For the off-diagonal elements there is
still the gauge freedom contained in $\cN_0$; choosing
$$ \cN_0 =\sqrt{ -{\Gamma^2(-\alpha)\Gamma(1-\gamma)\Gamma(1-\delta)
  \Gamma(\alpha+\gamma)\Gamma(\alpha+\delta) \over \Gamma^2(\alpha)
  \Gamma(\gamma)\Gamma(\delta)\Gamma(\beta+\gamma)\Gamma(\beta+\delta)}\,
  {}{d_1 \over d_2} }, \e$$
we arrive at
  $$  B = \sigma\, \pmatrix{ d & \rho \cr  \rho & -d^{-1} }, \e$$
with
$$\eqalign{
  d &:=\sqrt{d_1/d_2},\cr
  \sigma &:=\sqrt{ {s(\beta+\gamma)\,\s(\beta+\delta) \over s(\alpha)\,
  s(\beta) }\, {1 \over d_1d_2}}, \cr
  \rho &:=\sqrt{ - {s(\gamma)\,s(\delta) \over s(\beta+\gamma) \,
  s(\beta+\delta)} }. \cr}  \e$$

Using the above formula, eq.\ (59), yields the Boltzmann weights for any
IRF model based on some rational conformal field theory and a two block
primary field. In particular, for SU$(N)$ we recover the result derived
above, eq.\ (38). The calculation can be extended along the same lines to more
than two blocks. The three block case, which is applicable to many RCFT
(e.g., $C_n$, $B_n$, $D_n$ with the fundamental fields, which should recover
the height models previously described in ref. \r\Jimbo), has been considered,
and will be reported elsewhere
\REF\Future{J. Fuchs, work in progress}\r\Future.

We hope that the derivations described in this paper will be of help in the
general understanding of rational conformal field theory, along with
elucidating the connection with solvable lattice interaction round the face
models and soliton systems as described in ref. \r\I. Substituting
different RCFT in the general braid formulae described here, gives rise
to
a variety of new 
solvable lattice models which are of definite interest,
along with describing a host of new integrable field theories in two
dimensions.
\ack
The authors wish to thank the theory division of CERN for the
hospitability while part of this work was collaborated.
\refout

\end